\documentclass[aps,prx,twocolumn,superscriptaddress,showpacs,longbibliography]{revtex4-1}
\usepackage{amssymb,amsfonts,amsmath,amsthm}

\usepackage[dvips]{graphicx}
\usepackage{color}

\newcommand{\NN}{\mathbb{N}}
\newcommand{\RR}{\mathbb{R}}
\newcommand{\ZZ}{\mathbb{Z}}

\newtheorem{theo}{Theorem}
\newtheorem{coro}{Corollary}

\newtheorem{defi}{Definition}

\begin{document}

\title{Time-shift invariance determines the functional shape of the current
in dissipative rocking ratchets}

\author{Jos\'e A.\ Cuesta}
\email{cuesta@math.uc3m.es}
\affiliation{Grupo Interdisciplinar de Sistemas Complejos (GISC), Departamento
de Matem\'aticas, Universidad Carlos III de Madrid, Avda.~de la Universidad 30,
28911 Legan\'es, Spain}
\affiliation{Instituto de Biocomputaci\'on y F\'{\i}sica de Sistemas Complejos
(BIFI), Universidad de Zaragoza, 50009 Zaragoza, Spain}

\author{Niurka R.\ Quintero}
\email{niurka@us.es}
\affiliation{Instituto de Matem\'aticas de la Univesidad de Sevilla (IMUS), 41012 Sevilla, Spain}
\affiliation{Departamento de F\'\i sica Aplicada I, E.P.S., Universidad de
Sevilla, Virgen de \'Africa 7, 41011 Sevilla, Spain}

\author{Renato Alvarez-Nodarse}
\email{ran@us.es}
\affiliation{Instituto de Matem\'aticas de la Univesidad de Sevilla (IMUS), 41012 Sevilla, Spain}
\affiliation{Departamento de An\'alisis Matem\'atico, Universidad de Sevilla,
Apdo 1160, 41080 Sevilla, Spain}

\date{\today}

 
\begin{abstract} 
Ratchets are devices able to rectify an otherwise oscillatory behavior by
exploiting an asymmetry of the system. In rocking ratchets the asymmetry is
induced through a proper choice of external forces and modulations of nonlinear
symmetric potentials. The ratchet currents thus obtained in systems as different
as semiconductors, Josephson junctions, optical lattices, or ferrofluids, show
a set of universal features. A satisfactory explanation for them has challenged
theorist for decades, and so far we still lack a general theory of this
phenomenon. Here we provide such a theory by exploring ---through functional
analysis--- the constraints that the simple assumption of time-shift invariance of
the ratchet current imposes on its dependence on the external drivings. Because
the derivation is based on so general a principle, the resulting expression is
valid irrespective of the details and the nature of the physical systems to
which it is applied, and of whether they are classical, quantum, or stochastic.
The theory also explains deviations observed from universality under special
conditions, and allows us to make predictions of phenomena not yet observed in any
experiment or simulation.
\end{abstract}

%
%
\maketitle

\section{Introduction}

Forcing nonlinear transport systems with zero-average, time-periodic,
external forces may generate a ratchet current \cite{cole:2006}.
Ratchets are devices that exploit an asymmetry of the system (usually spatial)
to rectify an otherwise oscillatory behavior
\cite{zapata:1996,falo:1999,linke:1999,villegas:2003,beck:2005,falo:2002,costache:2010}.
The so-called rocking ratchets \cite{reimann:2002a,hanggi:2009} are able to do
so by breaking a temporal symmetry ---the external force cannot be reversed by a
time shift--- either in spatially symmetric systems
\cite{ajdari:1994} or in the presence of
some spatial asymmetry (see e.g. Refs. \cite{reimann:2002a,astumian:2002}). 
Ratchet currents can also be generated by a combined temporal and spatial symmetry
breaking \cite{poletti:2008,salger:2009}.

The two most studied mechanisms to induce a net current in a rocking
ratchet are \emph{harmonic mixing} \cite{reimann:2002a,hanggi:2009} and
\emph{gating} \cite{tarlie:1998,zamora-sillero:2006,gommers:2008}. In both of
them the involved periodic spatial potentials are symmetric. Harmonic mixing
amounts to imposing biharmonic external forces ---typically with a frequency
ratio 2:1--- and has been experimentally observed
\cite{schneider:1966,seeger:1978,schiavoni:2003,ustinov:2004,gommers:2005a,
ooi:2007,cubero:2010}
and theoretically studied
\cite{marchesoni:1986,goychuk:1998,flach:2000,morales-molina:2003}
in many different physical systems, both classical and quantum.
Biharmonic forces have also been used in experiments to modulate the
potential in some thermal ratchets devices
\cite{engel:2003,jager:2012}. In addition, harmonic mixing with more
than two harmonics has been explored in experiments with optical lattices
\cite{gommers:2006,gommers:2007}.

Gating ratchets also need at least two harmonics to break the temporal
symmetry, but they play a different role
\cite{zamora-sillero:2006,gommers:2008,noblin:2009}. In the most studied setup
one of the two harmonics acts as an external force whereas the other one is used
to modulate the spatial potential \cite{zamora-sillero:2006,gommers:2008}.   

Currents generated through many different rocking ratchets share a few
properties that hold regardless of the system. When two harmonics are used
and their amplitudes are small, the current exhibits a shifted sinusoidal shape
as a function of a precise combination of the phases of both harmonics. This
has been experimentally observed in semiconductors \cite{schneider:1966},
optical lattices \cite{schiavoni:2003}, ferrofluids \cite{engel:2003}, and
Josephson junctions \cite{ustinov:2004,ooi:2007} and has been theoretically
confirmed in studies of transport in semiconductors
\cite{schneider:1966}, Brownian particles
\cite{marchesoni:1986,wonneberger:1979}, solitons
\cite{zamora-sillero:2006,salerno:2002,morales-molina:2003}, ferrofluids
\cite{engel:2003}, and magnetic particles via dipolar interactions \cite{jager:2012},
among other systems. The phase lag of the sinusoid is known to depend on the
frequency of the harmonics, the damping, and other specific parameters of the
system \cite{breymayer:1984,borromeo:2005a} ---accordingly, current
reversals can be induced by acting on these parameters. Moreover, the ratchet
current is always found to be proportional to a product of specific powers of
the amplitudes of the harmonics.

Upon increasing the amplitudes of the harmonics beyond the small limit regime
departures from the sinusoidal behavior are observed, both in experiments
\cite{noblin:2009} and simulations \cite{cubero:2010}. As a consequence,
current reversals can also be induced by tuning the amplitudes of the harmonics
\cite{reimann:2002a,goychuk:1998}.
      
Although there have been many theoretical attempts to explain these universal
features of rocking ratchets, their scope is very limited, constrained to
specific models, and only applied to harmonic mixing. For instance, stochastic
theories have been used to explain the Brownian motion of a charged particle in
a periodic symmetric potential driven with a biharmonic force
\cite{wonneberger:1979,breymayer:1984}. Also, collective coordinate theories
have successfully explained harmonic mixing
\cite{salerno:2002,morales-molina:2003} and gating \cite{zamora-sillero:2006}
in soliton ratchets. For several models described by nonlinear differential
equations, symmetry properties of the current and of the systems can only
provide conditions on the two harmonics for a ratchet current to exist
\cite{reimann:2002a,ajdari:1994,zamora-sillero:2006,flach:2000}.

For decades, all attempts to reproduce the sinusoidal shape of the current have
failed to predict the existence of a system-dependent phase lag.
This lack of success is due to a flawed assumption ---widely employed in the
literature under the name of \emph{moment method}--- upon which all
these theories rely. According to this method, the ratchet current can
be obtained as an expansion in odd moments of the external force (starting
at the third moment because the time-average of the force is zero by
construction). That this method is generally incorrect has been shown
in Ref. \cite{quintero:2010} ---where the very restricted conditions for its
validity were properly delimited--- but it is easy to see why in an example:
If the force is a square wave all, its powers are proportional to the force
itself, and therefore the current must be zero. That this is not the case has
been shown in experiments \cite{arzola:2011}, simulations
\cite{ajdari:1994,schreier:1998}, and also theoretically in
Ref. \cite{quintero:2011}. The application of this method apparently captures the
right dependence on the amplitudes in the case of harmonic mixing, but this is
purely accidental. (For an in-depth analysis of this method and its many flaws,
see Refs. \cite{quintero:2010,quintero:2011} and references therein.)

An alternative theoretical approach has been recently proposed for the case
of harmonic mixing \cite{quintero:2010}. This theory does capture the nonzero
phase lag that the ratchet current normally exhibits and also predicts a
nonzero current for square-wave forces \cite{quintero:2011}. Nevertheless,
despite this relative success, a general theory that encompasses a unified
explanation of all universal features observed in so wide a diversity of
systems, an explanation of the deviations from them that occur outside the
small-amplitude regime and the effects induced by further harmonics, is still
lacking. Such a theory cannot be based on the particulars of specific systems
but has to rely on very general principles that hold for all of them.

In this paper, we explore the constraints that the simple time-shift invariance
satisfied by the ratchet current imposes on its shape and derive an expression
that explains all observations described above, both for harmonic mixing and
gating ratchets (with any number of harmonics). The formula describes correctly
not only the small-amplitude regime but also the deviations found for larger
amplitudes. And, because it is based on so general a principle, it is valid
regardless of the (dissipative) system and applicable even in the absence of a
mathematical model describing the phenomenon \cite{noblin:2009}. On top of
that, it allows us to make predictions so far not observed in any experiment or
simulation.

Before we enter into the details, a remark seems appropriate about what this
theory is \emph{not.} This theory is not meant to predict when a system does
exhibit a ratchet phenomenon. This is not possible because the theory
is so general that it holds both for dissipative systems that do and that do
not have ratchet currents. What the theory provides is a pattern to which any
ratchet current must conform. The theory claims that, under certain regularity
conditions, the ratchet current ---if any--- must necessarily be of a given
specific form. But the pattern depends on a set of unknown, system specific
coefficients that might all be zero ---hence yielding a zero current. For the
same reason the theory cannot predict any effect that depends on specific
details of the system. Having clarified this, what the theory does predict is
that the current must necessarily be zero if the system possesses some specific
symmetries ---so it is consistent with the well known fact that, unless some
symmetries are broken, a ratchet current cannot be generated
\cite{ajdari:1994,flach:2000,reimann:2002a}.

\section{General theory}

Suppose we have a physical system describing the position of a particle or
localized structure, $x(t)$, as a function of time. The system is driven by
some periodic, time-dependent, external driving $f(t)$ (external force,
parameter modulation, etc.). Function $x(t)$ ---or its expectation if the
system is stochastic--- is uniquely determined for any given $f(t)$, and so is
the ratchet current defined as \begin{equation}
v=\lim_{t\to\infty}\frac1t\int_0^t \dot{x}(\tau)d\tau
=\lim_{t\to\infty}\frac{x(t)-x(0)}t.  \end{equation} Mathematically this means
that the current $v$ is a functional of the external driving $f(t)$. Except for
very specific systems in which $v$ also depends on the initial conditions
(e.g., Hamiltonian systems or other nondissipative systems
\cite{salger:2009}), $v$ will ---by construction--- be invariant under time
shifts. We will show that the fact that $v$ is a time-invariant functional of
$f(t)$ is enough to determine the shape of the ratchet current for specific
drivings regardless of the system under study, as long as some regularity
assumptions of this functional dependence hold. Moreover, new symmetries of the
system can be incorporated into the theory to further specify this shape.

\subsection{Time-shift-invariant functionals of periodic functions}

Let $\mathcal{C}^s_T$, with $T>0$, be the set of continuous, $T$-periodic
functions $\mathbf{f}:\RR\to\RR^s$,
and let $\Gamma:\mathcal{C}_T^s \to\RR$ be a real functional on
$\mathcal{C}_T^s$. If $\Gamma$ is $n$ times Fr\'echet differentiable on
$\mathcal{C}_T^s$, then it has an $n$-th order Taylor expansion around
$\mathbf{0}$ \cite{wouk:1979}. Such a Taylor expansion can be
obtained as the $n$-th order truncation of the series \footnote{In Eq. 
\eqref{eq:Gamma}, it is implicitly assumed that if $n_i=0$ for some
$i=1,\dots,s$, variables $t_{ij}$ and factors $f(t_{ij})$ are missing; e.g.,
for $s=2$ and $\mathbf{n}=(n,0)$, the terms within the angular brackets
are $c_{n,0}(t_{11},\dots,t_{1n}) f_1(t_{11})\cdots f_1(t_{1n})$. With the
same convention, $c_{\mathbf{0}}$ is just a constant.}
\begin{equation}
\begin{split}
\Gamma[\mathbf{f}]=&\sum_{n_1=0}^{\infty}\cdots\sum_{n_s=0}^{\infty}
\langle c_{\mathbf{n}}(t_{11},\dots,t_{1n_1},\dots,t_{s1},\dots,t_{sn_s}) \\
&\times f_1(t_{11})\cdots f_1(t_{1n_1})\cdots
f_s(t_{s1})\cdots f_s(t_{sn_s})\rangle,
\end{split}
\label{eq:Gamma}
\end{equation}
where $\mathbf{n}=(n_1,\dots,n_s)$ and we have introduced the notation
\begin{equation}
\langle \Omega(t_1,\dots,t_r)\rangle=\frac{1}{T^r}\int_0^Tdt_1\cdots
\int_0^Tdt_r\,\Omega(t_1,\dots,t_r).
\end{equation}
The kernels $c_{n_1,\dots,n_s}(t_{11},\dots,t_{sn_s})$ are all real,
$T$-periodic, and symmetric in all their arguments.

In order to avoid cumbersome expressions we will henceforth work with the
full series \eqref{eq:Gamma}. It goes without saying that if $\Gamma$ is
at most $n$ times Fr\'echet differentiable the results we will obtain still
hold if the series are truncated at $n$th-order and an appropriate error
term is added \cite{wouk:1979}.

Consider the time-shift operator $(\mathcal{T}_{\tau}f)(t)=f(t+\tau)$. We
will say that $\Gamma$ is invariant under time shift if $\Gamma[\mathcal{T}_{\tau}
\mathbf{f}]=\Gamma[\mathbf{f}]$ for all $0<\tau<T$. Time-shift invariance
reflects on the kernels in Eq. \eqref{eq:Gamma} as the property
\begin{equation}
c_{n_1,\dots,n_s}(t_{11}-\tau,\dots,t_{sn_s}-\tau)=
c_{n_1,\dots,n_s}(t_{11},\dots,t_{sn_s})
\label{eq:ctshift}
\end{equation}
for all $0<\tau<T$.
\begin{theo}
\label{th:1}
Let $\Gamma$ be a time-shift-invariant functional with Taylor series
\eqref{eq:Gamma}, and take
\begin{equation}\label{f}
\mathbf{f}(t)=\big(\epsilon_1\cos(q_1\omega t+\phi_1),\dots,
\epsilon_s\cos(q_s\omega t+\phi_s)\big),
\end{equation}
where $\mathbf{q}\equiv(q_1,\dots,q_s)\in\NN^s$ is such that
$\gcd(q_1,\dots,q_s)=1$ \footnote{$\gcd(n_1,\dots,n_k)$ stands for ``greatest
common divisor'' of $n_1,\dots,n_k$, i.e., the largest integer that divides
all $n_1,\dots,n_k$.} and $\omega=2\pi/T$. Let $\mathcal{D}_+$
denote the set of nonzero solutions of the Diophantine equation
\footnote{The term Diophantine equation refers to an equation involving only
integer numbers. It is named after Diophantus of Alexandria, who introduced
them in his treatise \emph{Arithmetica}.}
$\mathbf{q}\cdot\mathbf{x}=q_1x_1+\cdots+q_sx_s=0$, whose leftmost nonzero
component is positive. Then,
\begin{equation}
\begin{split}
\Gamma[\mathbf{f}]=&\,
C_{\mathbf{0}}(\boldsymbol\epsilon) \\ &+
\sum_{\mathbf{x}\in\mathcal{D}_+}
\epsilon_1^{|x_1|}\cdots\epsilon_s^{|x_s|}
C_{\mathbf{x}}(\boldsymbol\epsilon)
\cos\big(\mathbf{x}\cdot\boldsymbol\phi+
\theta_{\mathbf{x}}(\boldsymbol\epsilon)\big),
\end{split}
\label{eq:cor1}
\end{equation}
where $\boldsymbol\phi\equiv(\phi_1,\dots,\phi_s)$, $\boldsymbol\epsilon\equiv(\epsilon_1,\dots,\epsilon_s)$, and functions
$C_{\mathbf{x}}(\boldsymbol\epsilon)$ and
$\theta_{\mathbf{x}}(\boldsymbol\epsilon)$ do not depend on $\boldsymbol\phi$
and are even in each $\epsilon_i$, $i=1,\dots,s$, for every
$\mathbf{x}\in\mathcal{D}_+$.
\end{theo}

(The proof of this theorem is deferred to Appendix~\ref{app:A}.)

When the functional $\Gamma$ exhibits further symmetries, some of the unknown
functions $C_{\mathbf{x}}(\boldsymbol\epsilon)$ and 
$\theta_{\mathbf{x}}(\boldsymbol\epsilon)$ in the expansion \eqref{eq:cor1} can
be determined. Two symmetries are important in this respect: \emph{force-reversal}
and \emph{time-reversal.}
\begin{defi}[Force reversal]
Let $I\subset\{1,\dots,s\}$ be a nonempty subset of indexes and let $\mathbf{f}:\RR
\to\RR^s$. We define the force-reversal operation $\mathcal{S}_{I}$ on $\mathbf{f}$
as the new vector function $\mathcal{S}_I\mathbf{f}(t)$ such that
$(\mathcal{S}_I\mathbf{f})_i(t)= -f_i(t)$ if $i\in I$ and
$(\mathcal{S}_I\mathbf{f})_i(t)= f_i(t)$ if $i\notin I$.
\end{defi}
\begin{coro}
\label{cor:2}
Under the conditions of Theorem~\ref{th:1}, let $I\subset\{1,\dots,s\}$
($I\ne\varnothing$). Then,
$\Gamma[\mathcal{S}_I\mathbf{f}]=-\Gamma[\mathbf{f}]$ if and only if
$C_{\mathbf{x}}(\boldsymbol\epsilon)=0$ for all
$\mathbf{x}\in\{\mathbf{0}\}\cup \mathcal{D}_+$ such that $\sum_{i\in I}x_i$ is
even.
\end{coro}
Since $C_{\mathbf{x}}(\boldsymbol\epsilon)$ is even in all its arguments, this
simply follows by replacing in Eq. \eqref{eq:cor1} $\epsilon_i$ by
$-\epsilon_i$ for all $i\in I$.
\begin{defi}[Time reversal]
Let $\mathbf{f}:\RR\to\RR^s$. We define the time-reversal operation $\mathcal{R}$ on
$\mathbf{f}(t)$ as $\mathcal{R}\mathbf{f}(t)=\mathbf{f}(-t)$.
\end{defi}
\begin{coro}
\label{cor:3}
Under the conditions of Theorem~\ref{th:1},
\begin{itemize}
\item[\rm (a)] $\Gamma[\mathcal{R}\mathbf{f}]=-\Gamma[\mathbf{f}]$ if and only if
$\theta_{\mathbf{x}}(\boldsymbol\epsilon)=\pm\pi/2$ for each $\mathbf{x}\in\mathcal{D}_+$ and
\item[\rm (b)] $\Gamma[\mathcal{R}\mathbf{f}]=\Gamma[\mathbf{f}]$ if and only if
$\theta_{\mathbf{x}}(\boldsymbol\epsilon)=0$ or $\pi$
for each $\mathbf{x}\in\mathcal{D}_+$.
\end{itemize}
\end{coro}
The proof of this corollary follows upon realizing that time-reversal amounts to
replacing $\phi_i$ by $-\phi_i$, for all $i=1,2,\ldots,s$, in \eqref{eq:cor1}.
 
\section{Application to different systems}

Equation~\eqref{eq:cor1} has been derived under the assumption that $\Gamma$ is a
sufficiently regular functional of $\mathbf{f}(t)$ and that it is time-shift
invariant.  Because these two assumptions are so general, it turns out that the
functional form \eqref{eq:cor1} must hold regardless of the specific system to which it 
is applied. In particular, details such as the kind of nonlinearities,
whether we deal with a particle or a localized field, the actual parameters,
etc., can only modify the functions $C_{\mathbf{x}}(\boldsymbol\epsilon)$ and
$\theta_{\mathbf{x}}(\boldsymbol\epsilon)$, and only in a very specific way
(they must be even functions of the amplitudes $\epsilon_j$). Furthermore, had
the system one of the symmetries of Corollaries~\ref{cor:2} and \ref{cor:3},
some of these functions would get automatically fixed regardless of any other
particular. This renders Eq. \eqref{eq:cor1} a universal expansion for the currents $v$
of rocking ratchets. In what follows, we discuss its application to explain different experimental
and numerical results reported in the literature of rocking ratchets.

\subsection{Two harmonic forces}

We start by considering systems for which the ratchet current arises from the
combined effect of two harmonics, $f_1(t)=\epsilon_1\cos(q\omega t+\phi_1)$
and $f_2(t)=\epsilon_2\cos(p\omega t+\phi_2)$. This special case is of great
importance because most rocking ratchets are induced by a biharmonic force
$f(t)=f_1(t)+f_2(t)$ \cite{schiavoni:2003,ustinov:2004,gommers:2005a,
ooi:2007,cubero:2010,marchesoni:1986,goychuk:1998,flach:2000,
morales-molina:2003,engel:2003}.
But it also comprises the so-called gating ratchets
\cite{hanggi:2009,zamora-sillero:2006,gommers:2008}, for which $f_1(t)$ is
an external force whereas $f_2(t)$ modulates the amplitude of a nonlinear
potential.

For two harmonics, the Diophantine equation $\mathbf{q}\cdot\mathbf{x}=0$
becomes $qx_1+px_2=0$. Its solutions are given by $\mathbf{x}=(kp,-kq)$,
$k\in\ZZ$, but those contributing to \eqref{eq:cor1} have
$k\in\NN_0(\equiv\NN\cup\{0\})$. Therefore $\mathbf{x}\cdot\boldsymbol\phi=
k(p\phi_1-q\phi_2)\equiv k\vartheta$, and \eqref{eq:cor1} reads
\begin{equation}
v[f_1,f_2]=\sum_{k=0}^\infty
(\epsilon_1^p\epsilon_2^q)^kC_k(\epsilon_1,\epsilon_2) 
\cos\big( k\vartheta+\theta_k(\epsilon_1,\epsilon_2)\big),
\label{eq:th1s=2}
\end{equation}
with $\theta_0(\epsilon_1,\epsilon_2)=0$.
 
Although Eq.~\eqref{eq:th1s=2} is valid for both, ratchets induced by a
biharmonic force and gating ratchets, their differences arise from their
different force-reversal symmetries. Let us analyze both cases separately.

\subsection{Ratchets induced by a biharmonic force}

In rocking ratchets with symmetric spatial potentials the current gets reversed
upon reversing the force (see,
e.g., Ref.~\cite{reimann:2002a,hanggi:2009,quintero:2010} and references therein).
Formally $v[-f_1,-f_2]=-v[f_1,f_2]$. Since $\gcd(p,q)=1$ either $p$ and $q$ are
both odd or have a different parity. In the former case $|x_1|+|x_2|=k(p+q)$ is
always even, so according to Corollary~\ref{th:1}, $C_k(\epsilon_1,\epsilon_2)=0$
for all $k\in\NN_0$ and therefore $v[f_1,f_2]=0$ (i.e., there is no ratchet
current). Notice that in this case $f(t+T/2)=-f(t)$, and since $v$ is 
time-shift invariant but changes sign under force reversal, it can only be
$0$. This explains our finding.

On the contrary, if $p+q$ is odd [in which case $f(t+\tau)\ne -f(t)$ for all
$\tau>0$], then Corollary~\ref{th:1} implies only
$C_{2k}(\epsilon_1,\epsilon_2)=0$, $k\in\NN$; hence
\begin{equation}
\begin{split}
v=& \sum_{\substack{k=1 \\ k\text{ odd}}}^{\infty}
(\epsilon_1^p\epsilon_2^q)^kC_k(\epsilon_1,\epsilon_2)
\cos\big(k\vartheta+\theta_k(\epsilon_1,\epsilon_2)\big).
\end{split}
\label{eq:th1s=2fr}
\end{equation}
The lowest order in Eq. \eqref{eq:th1s=2fr} yields
\begin{equation}
v=C_1(0,0)\epsilon_1^p\epsilon_2^q
\cos\big(\vartheta+\theta_1(0,0)\big)+o(\epsilon_1^p\epsilon_2^q)
\label{eq:th1s=2fr-shape}
\end{equation}
a result first obtained in Eq. \cite{quintero:2010}.

But, Eq.~\eqref{eq:th1s=2fr} contains more information. The lowest order
at which the next harmonic enters in $v$ is $O(\epsilon_1^p\epsilon_2^q)^3$.
For the simplest ---and most common--- case studied in the literature, namely,
$p=2$ and $q=1$, this implies that the second harmonic first appears at ninth
order. Therefore, an improvement on Eq. \eqref{eq:th1s=2fr-shape} is
\begin{equation}
v=C_1(\epsilon_1,\epsilon_2)\epsilon_1^2\epsilon_2
\cos\big(\vartheta+\theta_1(\epsilon_1,\epsilon_2)\big)
+E_9(\epsilon_1,\epsilon_2),
\label{eq:th1s=2fr-shape-amplitudes}
\end{equation}
where the error $E_9(\epsilon_1,\epsilon_2)$ contains terms of order 9 or
higher, and $C_1(\epsilon_1,\epsilon_2)$ and $\theta_1(\epsilon_1,\epsilon_2)$
are  quadratic polynomials in $\epsilon_1^2$ and $\epsilon_2^2$.
Equation~\eqref{eq:th1s=2fr-shape-amplitudes} tells us that, whereas Eq. 
\eqref{eq:th1s=2fr-shape} captures the shape of the ratchet current for
sufficiently small amplitudes, upon increasing the amplitudes we can modify the
phase lag $\theta_1(\epsilon_1,\epsilon_2)$. Put in a different way, if we fix the
phases $\phi_1$ and $\phi_2$ of the biharmonic force so that
$\vartheta=-\theta_1(0,0)+\pi/2$, the ratchet current is suppressed
\cite{gommers:2005a,borromeo:2005a}. But then we can restore it \emph{without
changing the phases} by increasing the amplitudes. 

This current reversal was observed in experiments
\cite{gommers:2005a,cubero:2010} and attributed to a dissipation-induced
symmetry breaking. Our Eq.~\eqref{eq:th1s=2fr-shape-amplitudes} reveals that
this is the default behavior of a ratchet like this, because the current
vanishes at a value of $\vartheta$ that depends not only on the amplitudes of
the biharmonic force, but also on the frequency and other parameters of the
system.

\begin{figure}
\centerline{\includegraphics[width=92mm,clip=]{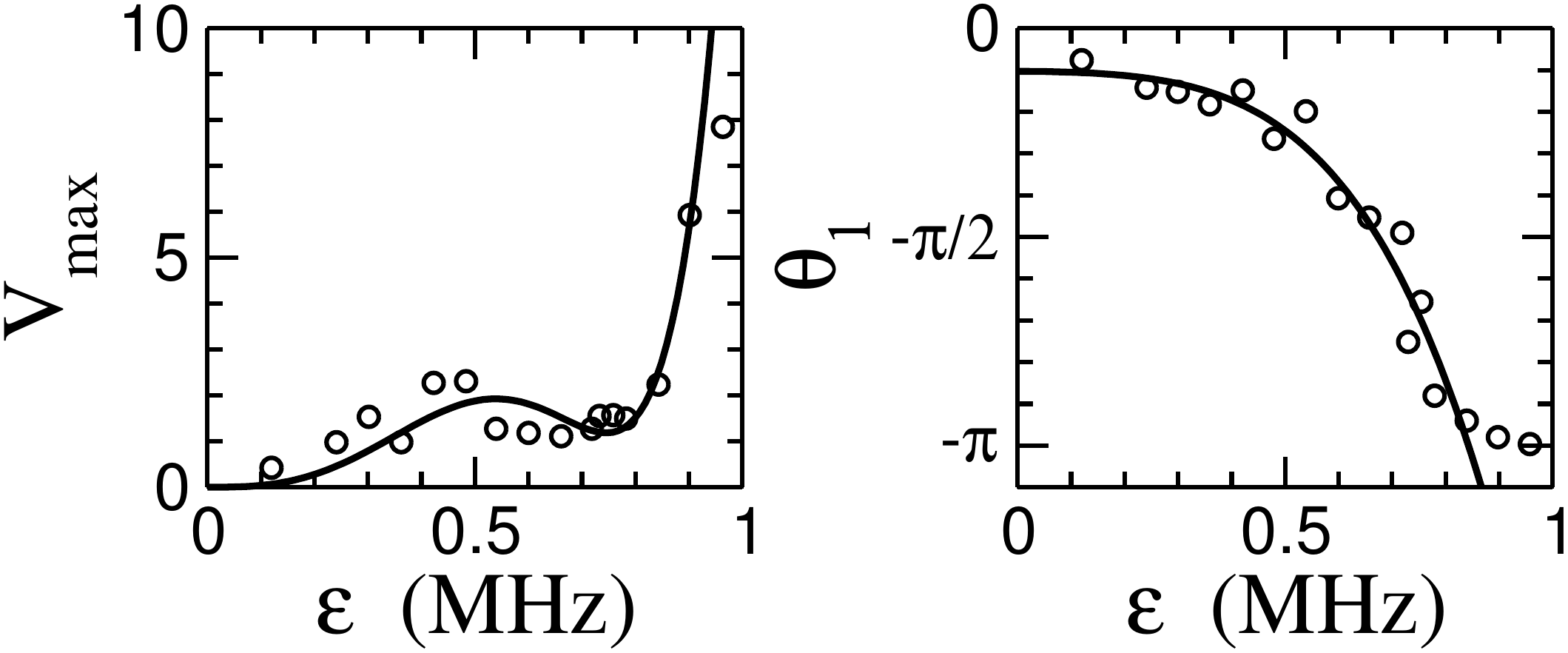}}
\caption{Maximum ratchet velocity $v_{\mbox{\tiny max}}$ and phase
lag $\theta_1$ as functions of the amplitude
$\epsilon=\epsilon_1=\epsilon_2$ of the biharmonic force, for the rocking
ratchet of Ref.~\cite{cubero:2010}. Points are the experimental data, and lines
are fits to formula~\eqref{eq:th1s=2fr-shape-amplitudes}:
$\theta_1(\epsilon)=-0.3238-0.5996\epsilon^2-4.7445 \epsilon^4$ and
$v_{\text{max}}(\epsilon)=\epsilon^3(39.631-124.661\epsilon^2
+105.258\epsilon^4)$.
\label{fig1}}
\end{figure}

Functions $C_{1}(\epsilon_{1},\epsilon_{2})$ and
$\theta_{1}(\epsilon_{1},\epsilon_{2})$ are experimentally obtained for
a range of values of $\epsilon_1=\epsilon_2=\epsilon$ \cite{cubero:2010}. 
Figure~\ref{fig1} shows a fit of the experimental data to the curve
$v=C_1(\epsilon)\epsilon^3\cos\big(\vartheta+\theta_1(\epsilon)\big)$, with
$v_{max}=C_{1}(\epsilon) \epsilon^3$ being $C_1(\epsilon)$ and
$\theta_1(\epsilon)$ quadratic polynomials in $\epsilon^2$.

Another prediction of the theory follows from Corollary~\ref{cor:3}: For
systems having either of those two symmetries upon time reversal, all
phase lags $\theta_k(\boldsymbol\epsilon)$ in the expansion \eqref{eq:th1s=2fr}
are constant ---either $0$ or $\pi$, or $\pm\pi/2$, depending on the
symmetry. This is confirmed, e.g., by simulations carried out on the Langevin
equation
\begin{equation}
\alpha\dot x=U_0k\sin(2kx)+f(t)+\eta(t),
\label{eq:Langevin}
\end{equation}
with $\eta(t)$ a zero-mean white noise such that $\langle\eta(t)\eta(t')\rangle
=D\delta(t-t')$ and $f(t)$ is a biharmonic force \cite{cubero:2010}.
Figure~4 (upper panel) of Ref.~\cite{cubero:2010} shows that
$v(\pm\pi/2)=0$ for all amplitudes. (In this overdamped regime, the velocity
does not change sign upon time reversal.) This figure is especially revealing because
for the largest amplitudes, the velocity clearly shows the influence of the
second harmonic, and yet the phase lags remain constant.


\subsection{Gating ratchets}

Force reversal acts differently for gating ratchets because, of the two
harmonics, only $f_1(t)$ is an external force. In this case,
when the potential is symmetric \cite{hanggi:2009,zamora-sillero:2006,gommers:2008},
we have $v[-f_1,f_2]=-v[f_1,f_2]$. Thus, Corollary~\ref{th:1} implies
$C_k(\epsilon_1,\epsilon_2)=0$ if $kp$ is even ($k\in\NN_0$). If $p$ is even,
then $v=0$, whereas if $p$ is odd, then only $C_{2k}(\epsilon_1,\epsilon_2)=0$,
$k\in\NN$, and we again recover Eqs. \eqref{eq:th1s=2fr} and
\eqref{eq:th1s=2fr-shape}. Notice that if $p$ is even, then $q$ must be odd
[because $\gcd(p,q)=1$], and therefore $f_1(t+T/2)=-f_1(t)$ and
$f_2(t+T/2)=f_2(t)$. Thus, a time shift can reverse the current ---which means
that the current must be zero.

Thus, the ratchet currents produced by either gating or a biharmonic force are
both given by the same formula. There is an exception, though: Gating does not
put any constraint on $q$, so a ratchet current can be obtained even for $q=p=1$
\cite{hanggi:2009,zamora-sillero:2006,gommers:2008}. For this particular
case, the lowest order at which the second harmonic shows up in the current is
the sixth, i.e.,
\begin{equation}
v=C_1(\epsilon_1,\epsilon_2)\epsilon_1\epsilon_2
\cos\big(\vartheta+\theta_1(\epsilon_1,\epsilon_2)\big)
+E_6(\epsilon_1,\epsilon_2),
\label{eq:gaing-amplitude}
\end{equation}
and $C_1(\epsilon_1,\epsilon_2)$ and $\theta_1(\epsilon_1,\epsilon_2)$ are
linear in $\epsilon_1^2$ and $\epsilon_2^2$. Accordingly, a shift of the phase
lag with the amplitudes similar to that observed in biharmonic ratchets
\cite{cubero:2010} is to be expected in gating ratchets. Thus, not only has
formula \eqref{eq:gaing-amplitude} been obtained here for the first
time (to the best of our knowledge, no theory has ever been attempted to
explain the current observed in gating ratchets) but the possibility of
inverting the current by varying the amplitudes of the harmonics in these
systems is a prediction of this theory that, as far as we know, still needs
experimental confirmation.

\subsection{Particles moving in asymmetric potentials}

An interesting case to analyze with the theory is that of particles moving (or
solitons lying) in potentials lacking mirror symmetry. In these cases, the
current does not have the force-reversal symmetry exploited above because the mirror image of the
system is a different system. Then, all terms in \eqref{eq:th1s=2} are nonzero
in principle. In the case of two harmonics ---irrespective of whether we are
considering ratchets induced by biharmonic forces or gating ratchets--- the
lowest order in the expansion \eqref{eq:th1s=2} is given by
$C_0(\epsilon_1,\epsilon_2)$, a polynomial of $\epsilon_1^2$ and
$\epsilon_2^2$. Clearly, $C_0(0,0)=0$ if there is no ratchet current in
the absence of external force; therefore, in this case, the theory predicts,
for small amplitudes, a ratchet current \emph{independent of the phases}
(a dependence that may be restored at higher orders) and proportional to a
linear combination of $\epsilon_1^2$ and $\epsilon_2^2$.

As a matter of fact, the theory also predicts that \emph{even with a single
harmonic} (say, $\epsilon_2=0$), a ratchet current proportional to $\epsilon_1^2$
can be generated.  This is indeed what was found in Refs.
\cite{reimann:2002a,quintero:2005a}. In this case we also know from Eq.
\eqref{eq:th1s=2} that all higher-order terms are identically zero, so the
prediction is even stronger: The current must be of the form
$\epsilon_1^2Q(\epsilon_1^2)$, with $Q(x)$ a certain function.  Notice in
particular that, depending on whether $Q(x)$ does or does not change sign, the
current may or may not exhibit reversals upon variations of the amplitude
$\epsilon_1$.

\subsection{Other ratchets with two harmonics}

\begin{figure}
\begin{center}
\includegraphics[width=70mm]{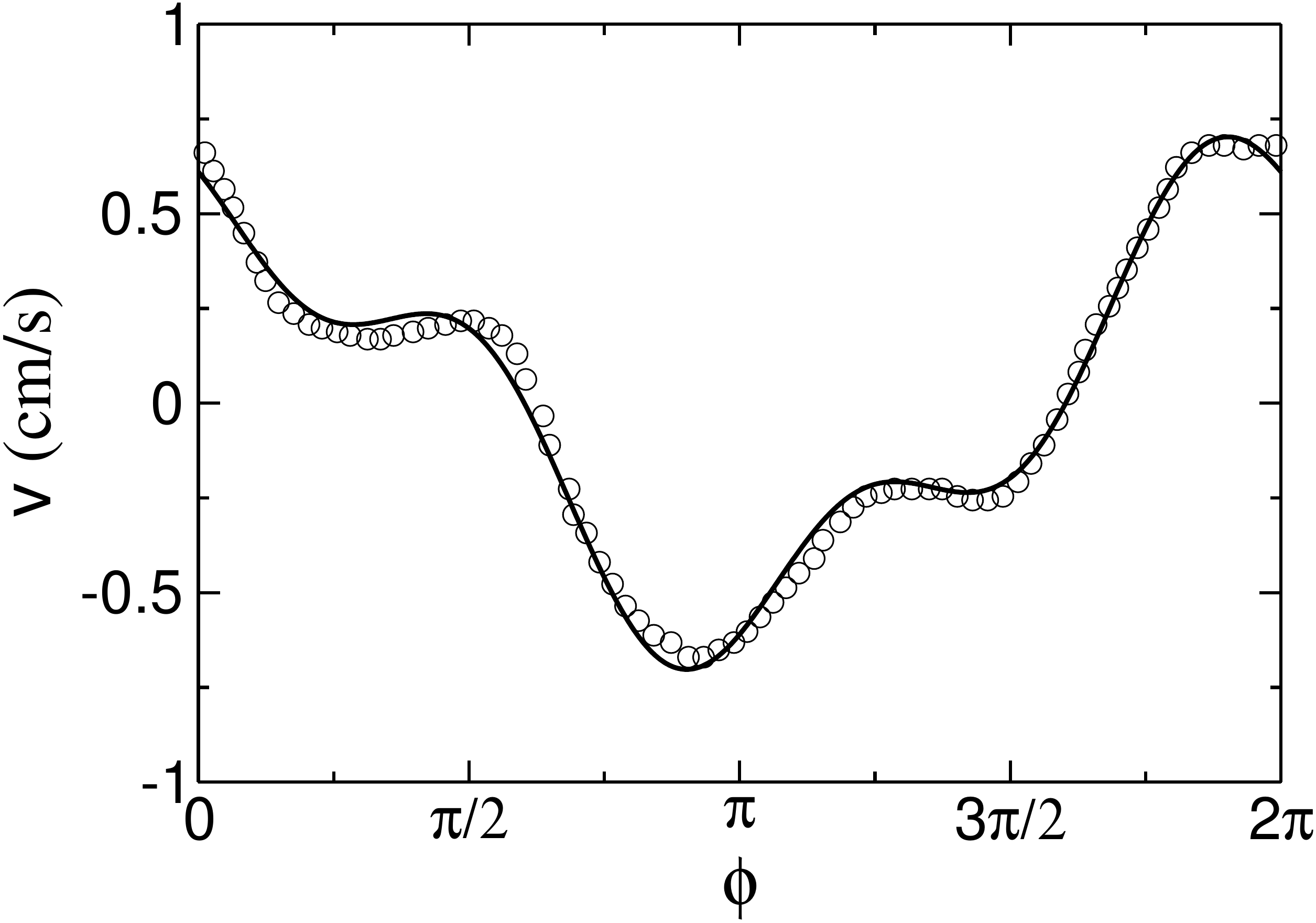}
\end{center} 
caption{Droplet velocity $v$ as function of the phase shift $\phi$ between the
horizontal and vertical vibrations of the plate. Symbols represent experimental data
from Figure~3, upper panel, $A_h=240\mu{m}$, of Ref.~\cite{noblin:2009}. The line represents the
fit of the curve $v(\phi)=0.5435\cos(\phi-0.0211)+0.1972\cos(3\phi+1.2267)$.
\label{fig-nob-pnas}}
\end{figure}

Liquid drops on a horizontal plate exhibit ratchet movement when the plate
is vibrated with both horizontal and vertical harmonic forces
\cite{noblin:2009}. These forces have the same frequency and a relative phase
$\phi$, and as usual the ratchet current depends on $\phi$. We are not aware of
any theoretical approach that explains why the average velocity $v$ of the
drops exhibits a nonsinusoidal behavior as a function of the relative phase
shift $\phi$. However, Fig.~3(a) of Ref.~\cite{noblin:2009} reveals that $v$
changes sign when the vertical force $f_1$ is reversed; i.e.,
$v[-f_1,f_2]=-v[f_1,f_2]$. According to our approach, this is enough to
conclude that the drop velocity must behave as the current of a gating ratchet.
Hence, it will be given by Eq.~\eqref{eq:th1s=2fr} for $p=q=1$.
Figure~\ref{fig-nob-pnas} shows a fit with the first two harmonics of this
equation to the experimental data of Ref.~\cite{noblin:2009}. 

This anharmonicity is also predicted by our theory when the ratchet is induced
by a biharmonic force with large amplitudes, and it has been reported recently
in simulations of classical particles in a one-dimensional driven superlattice
\cite{wulf:2012}.

\subsection{Forcing with more than two harmonics}

In some experiments with cold atoms \cite{gommers:2006,gommers:2007}, ratchets
are generated using more than two harmonics. The simplest one is of the form
\begin{equation}
f(t)=a\big[\cos(q\omega t+\phi_1)+\cos(2q\omega t+\phi_2)+
\cos(p\omega t+\phi_3)\big].
\label{eq:tri-harmonic}
\end{equation}
Although the Diophantine equation $qx_1+2qx_2+px_3=0$ has three unknowns,
the solution can be readily obtained using Blankinship's algorithm
\cite{morito:1980} as
$\mathbf{x}=k_1\mathbf{s}_1+k_2\mathbf{s}_2$, where $\mathbf{s}_1=(2,-1,0)$,
$\mathbf{s}_2=(p,0,-q)$, and $k_1,k_2\in\ZZ$. Hence, $\mathbf{x}=
(2k_1+pk_2,-k_1,-qk_2)$. The subset contributing to \eqref{eq:cor1} is defined
by $2k_1+pk_2\in\NN_0$; on the other hand, because of force reversal
(c.f.~Corollary~\ref{cor:2}), the only nonzero coefficients have
$3k_1+(p+q)k_2$ odd. Hence, if $p+q$ is even, then $k_1$ must be odd, whereas
if $p+q$ is odd, then $k_1+k_2$ must be odd. Then,
\begin{equation}
\begin{split}
v= &\sum_{\mathbf{k}\in\Omega_{p,q}}C_{\mathbf{k}}(a)
a^{2k_1+pk_2+|k_1|+q|k_2|} \\
&\times
\cos\big(k_1\vartheta_1+k_2\vartheta_2+\theta_{\mathbf{k}}(a)\big),
\end{split}
\label{eq:triharmonic-even}
\end{equation}
where $\vartheta_1=2\phi_1-\phi_2$, $\vartheta_2=p\phi_1-q\phi_3$, and
$\Omega_{p,q}=\{(k_1,k_2)\in\ZZ^2\,:\,2k_1+pk_2\ge 0, \, k_1\text{ odd}\}$ if
$p+q$ is even, or $\Omega_{p,q}=\{(k_1,k_2)\in\ZZ^2\,:\,2k_1+pk_2\ge 0, \,
k_1+k_2\text{ odd}\}$ if $p+q$ is odd.

The choice $q=p=1$ \cite{gommers:2006} reduces Eq. \eqref{eq:tri-harmonic} to a
biharmonic force where the amplitude of the one of harmonics depends on the
phase $\phi_3$.  In other words, the shape of the current is again a sinusoidal
function of $\vartheta_1$, with the usual cubic prefactor of the amplitudes;
however, in this case both the maximum current and the phase lag depend on
$\phi_3$. This is exactly what the experiments reveal (c.f.~Fig.~1 of
Ref.~\cite{gommers:2006}).

Another relevant choice of parameters is $q,p\to\infty$ and $q\not=p$
\cite{gommers:2006}. For $a<1$, it implies that $k_2=0$. Thus, regardless of the
parity of $p+q$, Eq. \eqref{eq:triharmonic-even} becomes
\begin{equation}
\begin{split}
v =& \sum_{\substack{k=1 \\ k\text{ odd}}}^{\infty}C_{k,0}(a)a^{3k}
\cos\big[k\vartheta_1+\theta_{k,0}(a)\big].
\end{split}
\end{equation}
The lowest order is
$v=C_{1,0}(0)a^3\cos\big(\vartheta_1+\theta_{1,0}(0)\big)+O(a^5)$,
which explains the observations made in Ref. \cite{gommers:2006}, namely, the
sinusoidal dependence on $\vartheta_1$ and the insensitivity of $\theta_{1,0}(0)$
to variations of the phase $\phi_3$. 

The limit case $p,q\to\infty$ is particularly interesting because it connects
the effect of perturbations with quasiperiodic forces. Suppose the
harmonics depend on two frequencies, $\omega_1$ and $\omega_2$, such that
$\omega_2/\omega_1$ is not a rational number. One can choose rational
approximants $p/q$ of $\omega_2/\omega_1$ such that $\omega_1\approx q\omega$
and $\omega_2\approx p\omega$ for a suitable $\omega$. The theory can then be
applied for each choice of $p$ and $q$, and the quasiperiodic limit can be
recovered as the limit $p,q\to\infty$ and $\omega\to 0$ with
$p/q\to\omega_2/\omega_1$. For an illustration of the application of this
method, we refer to the appendix of Ref.~\cite{cubero:2012}.

A second more complicated forcing has also been tested for cold atoms
\cite{gommers:2007}. The force in this case can be cast
as a sum of four harmonics $f(t)=f_1(t)+f_2(t)+f_3(t)+f_4(t)$, where
\begin{equation}
\begin{split}
f_1(t)=& \frac{b}{2}(2q+p)\cos\left((2q+p)\omega t+2\phi_1+\phi_2\right), \\
f_2(t)=& \frac{b}{2}(2q-p)\cos\left((2q-p)\omega t+2\phi_1-\phi_2\right), \\
f_3(t)=& \frac{a}{2}(q+p)\cos\left((q+p)\omega t+\phi_1+\phi_2\right), \\
f_4(t)=& \frac{a}{2}(q-p)\cos\left((q-p)\omega t+\phi_1-\phi_2\right).
\end{split}
\label{eq:4harmonic}
\end{equation}
 Two cases have been studied \cite{gommers:2007}:
$q=p=1$ and $q=3$, $p=2$.

For $q=p=1$, $f_4(t)=0$, and there are three harmonics left. The expansion of $v$
in terms of $\vartheta=\phi_2-\phi_1$ and the amplitudes can be obtained
using a similar procedure [see Appendix~\ref{app:B}, Eq.~\eqref{eq:v4h3}]. To
lowest order, 
\begin{equation}
\begin{split}
v= &b^2a\big\{C\cos(3\vartheta+\theta_0)
+D\cos(\vartheta+\theta_1)\big\}+E_5(a,b),
\end{split}
\label{eq:4h3}
\end{equation}
where $E_5(a,b)$ contains five-order terms in $a$ and $b$. This expression
features, even at the lowest order in the amplitudes, a deviation from the
usual sinusoidal shape. In the experiments, the second harmonic went unnoticed
because at that time no available theory predicted any such deviation.
However, the fit of the experimental data to a cosine function shows a
systematic discrepancy that might be the fingerprint of this second harmonic
(c.f.~Fig.~1 of Ref. \cite{gommers:2007}). Further experiments should reveal this
second harmonic more clearly.

The second case experimentally tested is $q=3$, $p=2$. For this case,
all four harmonics \eqref{eq:4harmonic} are present. The full expansion
in terms of $\vartheta=3\phi_2-2\phi_1$ and the amplitudes is obtained
in Appendix~\ref{app:B} [c.f.~Eq.~\eqref{eq:v4h4}]. To lowest order,
\begin{equation}
v=A(a,b)\cos\big(\vartheta+\psi(a,b)\big)+E_5(a,b),
\label{eq:phaseamplitude}
\end{equation}
with $A(a,b)$ and $\psi(a,b)$ given by \eqref{lasteq}.

The usual cosine shape of the current was already observed in the experiments
\cite{gommers:2007}. However, Eq.~\eqref{eq:phaseamplitude} reveals an
unexpected new effect. In harmonic mixing currents, it is customary to set
$a=r\epsilon$ and $b=(1-r)\epsilon$ and vary $0\le r\le 1$. If the system
is driven by a biharmonic force, changing $r$ changes the intensity of the
current \cite{schiavoni:2003}. However, if $\epsilon$ is sufficiently small,
the phase at which the current vanishes does not depend on $r$. In other words,
if $\vartheta$ is fixed to this phase and $r$ is varied, no ratchet current is
produced. (As explained before [c.f.~Eq.~\eqref{eq:th1s=2fr-shape-amplitudes}],
this is no longer true if $\epsilon$ is large; see also
Ref.~\cite{cubero:2010}.) But Eq.~\eqref{eq:phaseamplitude} tells us
that $\psi(a,b)$ does depend on $r$ \emph{even for small $\epsilon$.} This
implies that by setting $\vartheta$ so that the current is zero for a given
$r$, we can generate a ratchet current by simply changing $r$.

To confirm this prediction of the theory we carry out simulations
for the damped sine-Gordon equation
\begin{equation}
\psi_{tt}-\psi_{xx} + \sin\psi + \beta \psi_t = f(t)
\label{eq:phi}
\end{equation}
driven by the multifrequency force
\eqref{eq:4harmonic} with $q=3$, $p=2$. We have solved numerically this
equation in the interval $[-70,70]$, with periodic boundary conditions, by
discretizing the second spatial derivative using centered finite differences on
a grid of step size $\Delta x=0.1$. We have integrated the resulting set of
ordinary differential equations with a fourth-order Runge-Kutta method along 10
complete periods, with a time step $\Delta t=0.01$.  As the initial condition,
we use an exact static one-soliton solution, centered at zero, of the
unforced [$f(t)=0$] and undamped ($\beta=0$) sine-Gordon
equation~\eqref{eq:phi}.

\begin{figure}[t]
\begin{center}
\includegraphics[width=70mm]{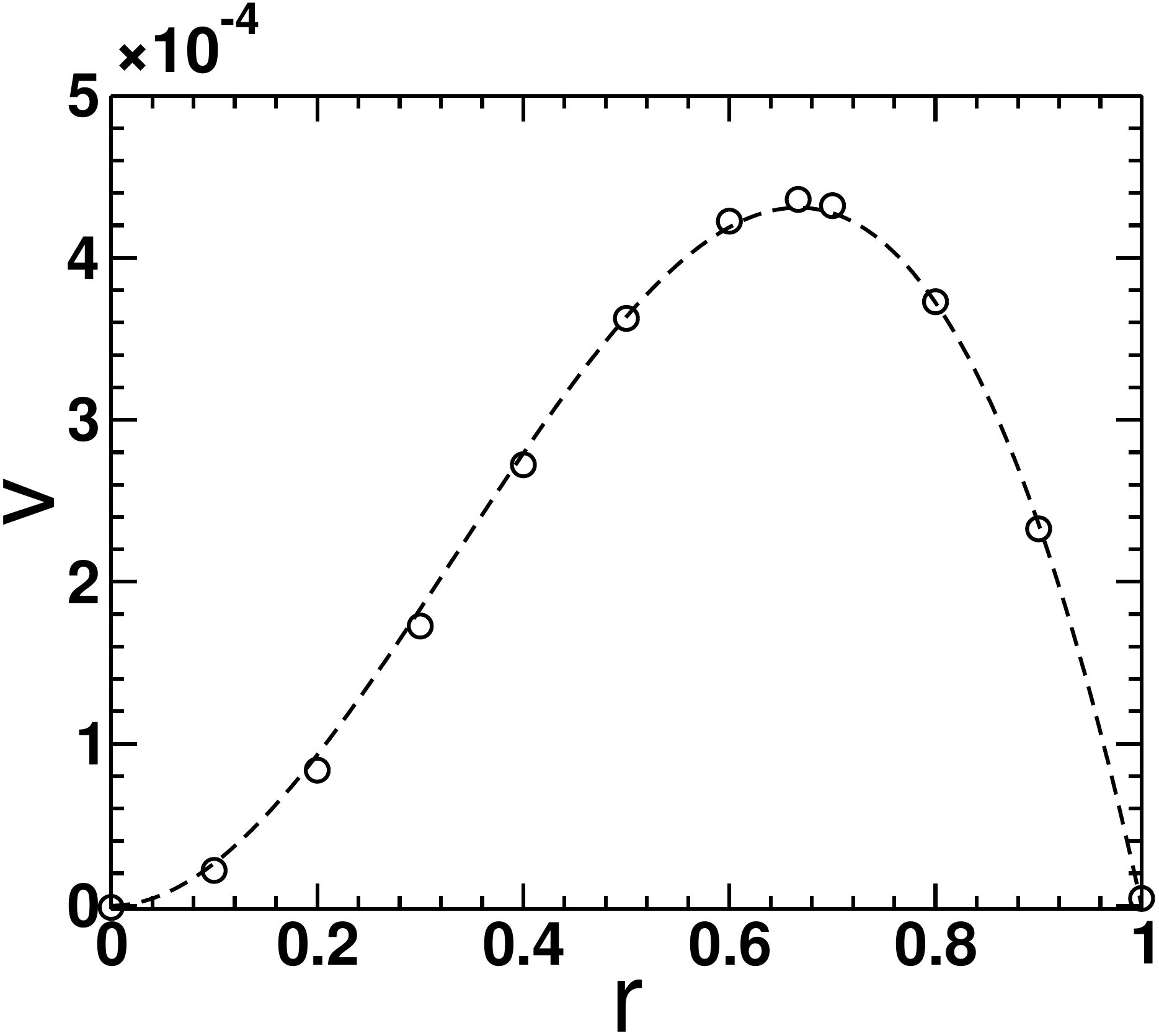}
\end{center}
\caption{Ratchet effect induced in a damped sine-Gordon system
(c.f.~\eqref{eq:phi}) driven by the force \eqref{eq:4harmonic} upon changing
the amplitudes $a=r\epsilon$ and $b=(1-r)\epsilon$. The velocity $v$ is plotted
as a function of $r$. Parameters are $\omega=0.05$, $\phi_1=\pi/2$,
$\beta=0.05$, and $\epsilon=0.03$. Phase $\phi_2=0.64486$ is chosen so
that $v$ vanishes for $r=0$. Circles are results obtained from the simulations.
Line is the fitted curve $v=0.0029089 (1-r) r^2$.
\label{fig-vr} 
}
\end{figure}

Notice that if $r=0$ or $r=1$, only two of the four harmonics
\eqref{eq:4harmonic} remain. Their frequencies are such that for $r=1$,
$f(t)=-f(t+T/2)$ ---hence, there is no ratchet current---
whereas for $r=0$ this symmetry is broken ---hence there is a ratchet
current. Accordingly, we set $r=0$ and $\phi_1=\pi/2$, and find the value of
$\phi_2$ for which $v=0$. Then, we fix this value for $\phi_2$ and vary
$r$. The result is shown in Fig.~\ref{fig-vr}. As predicted, varying
$r$ induces a ratchet current. As a matter of fact, the numerical values
fit perfectly the theoretical prediction $v\propto\epsilon^3(1-r)r^2$
that follows from Eqs. \eqref{eq:phaseamplitude} and \eqref{blasteq}.

\section{Discussion and conclusions} 

In this work, we have introduced a theory that captures, in a unified framework,
the ratchet transport generated by zero-average, periodic drivings of very
different kinds of systems, like cold atoms in optical lattices, fluxons in
Josephson junctions, current in semiconductors, or transport of ferromagnetic
nanoparticles in liquids. The theory can be applied to classical or quantum
dissipative
systems alike, with or without thermal fluctuations. The number of different
harmonics the theory can deal with is arbitrary.  Although most studies use
two, added up in a single biharmonic force or used for two different purposes
(like a force and a potential modulation
\cite{zamora-sillero:2006,gommers:2008} or two independent forces
\cite{noblin:2009}), the theory also explains experiments carried out driving
the system with three or four harmonics \cite{gommers:2006,gommers:2007}, as
well as experiments in ferrofluids, where the biharmonic force modulates the
potential in a new type of thermal ratchet device \cite{engel:2003}. 

Focusing on the results for two harmonics, Eq.~\eqref{eq:th1s=2} already
captures many universal features observed in experiments and simulations.
First, it shows the widespread sinusoidal dependence observed when the
amplitude of the external forces is small
\cite{schneider:1966,zamora-sillero:2006,gommers:2008,schiavoni:2003,
ustinov:2004,ooi:2007,goychuk:1998,morales-molina:2003,engel:2003}.
Second, it explains why the sinusoid is observed even when the amplitude of the
force is not so small \cite{cubero:2010}. Third, it captures the departures of
this sinusoidal shape for even larger amplitudes
\cite{cubero:2010,noblin:2009}. And fourth, it shows that the point where the
current vanishes (the phase lag) depends on the amplitude, the frequency, and
the rest of the system parameters. In particular, this means that we can
generate or revert the current by simply changing the amplitudes of the two
harmonics \cite{cubero:2010,noblin:2009}, their frequency
\cite{breymayer:1984,gommers:2005b}, or (rather paradoxically) the damping in
systems with dissipation
\cite{gommers:2005a,breymayer:1984,morales-molina:2006,quintero:2011}. If the
system satisfies certain symmetries, the theory predicts that the phase lags can
no longer be modified by changing the amplitudes of the harmonics
(Corollary~\ref{cor:3}). This is indeed what happens in some equations for
particles or solitons moving in a nonlinear potential and in certain
experiments \cite{ooi:2007,marchesoni:1986,engel:2003,quintero:2011}.

One of the most remarkable facts about this theory is its universality. In
its derivation, we have simply used  two assumptions: (a) The velocity is a
sufficiently regular functional of the external force (regularity condition),
and (b) it is invariant under time shifts (time-shift symmetry). The former is
used to make a Taylor expansion ---perhaps only up to some finite order--- of
the velocity with respect to the external force; the latter leads, in the case
of harmonic forcings, to a Fourier expansion in terms of some combination of
the phase shifts between the harmonics. The fact that the functional form
\eqref{eq:cor1} is obtained under so general assumptions implies that the
particulars of the system under study (e.g., the kind of nonlinearities or the
specific parameters) can only tune the constants but never change the
functional form. As a matter of fact, we do not even need to have an explicit
mathematical model of the experimental system to predict how the velocity
depends on the phases of the harmonics and to constrain its dependence on the
amplitudes (e.g., the case analyzed in Fig.~\ref{fig-nob-pnas}).

Of the two assumptions above, only regularity limits the applicability of the
theory. Besides, it might be a requirement that is hard to verify for a given physical
system. Nonetheless, the success of the theory in explaining the results of so
many different experimental and numerical sources suggests that the systems to
which it does not apply must be rare.
Exceptions can be found, though. For instance, simulations of the discrete
Frenkel-Kontorova system show discontinuities in the behavior of the current
as a function of the phases in the biharmonic force \cite{zolotaryuk:2006}. 
Also, the ratchet current of periodically forced overdamped particles moving in
an asymmetric potential exhibits discontinuities as a function of the amplitude
of the forcing \cite{bartussek:1994,ajdari:1994} ---although these
discontinuities disappear in the presence of noise, which may thus be acting as
a regularizer of the functional.

We end by pointing out that universality permits us not only to
explain a plethora of specific phenomena or anomalies that different
experiments and simulations have evidenced but also to predict new ones that
have not been observed yet and need experimental confirmation. Some of them are
described above, and some others have been stated along the way while
analyzing systems which had been experimentally studied. But, by making
specific choices for the number of harmonics and their frequencies in
Eq.~\eqref{eq:cor1}, many more can be derived.

\bibstyle{revtex}
\bibliography{ratchets}

\appendix
\section{Proof of Theorem~\ref{th:1}}
\label{app:A}

Writing the cosines as complex exponentials and substituting
in Eq. \eqref{eq:Gamma} leads to
\begin{equation}
\begin{split}
\Gamma[\mathbf{f}]=&\sum_{\mathbf{k},\mathbf{l}\in\NN_0^s}
\epsilon_1^{k_1+l_1}\cdots \epsilon_s^{k_s+l_s}
A(\mathbf{k},\mathbf{l})e^{i(\mathbf{k}-\mathbf{l})\cdot\boldsymbol\phi},
\end{split}
\label{eq:Gammaf}
\end{equation}
where
\begin{equation}
\begin{split}
A(\mathbf{k},\mathbf{l}) =&\prod_{j=1}^s\frac{(k_j+l_j)!}{2^{k_j+l_j}
k_j!l_j!} \\
&\times
\widehat c_{\mathbf{k}+\mathbf{l}}(\{q_1\}_{k_1},\{-q_1\}_{l_1},\dots,
\{q_s\}_{k_s},\{-q_s\}_{l_s}).
\end{split}
\label{eq:A}
\end{equation}
Here, we are using the shorthand
$\{a\}_k=\overbrace{a,\dots,a}^{k\text{ times}}$
and denoting
\begin{equation}
\widehat c_{\mathbf{n}}(r_1,\dots,r_{|\mathbf{n}|})=
\left\langle c_{\mathbf{n}}(t_1,\dots,t_{|\mathbf{n}|})e^{i\omega
\left(r_1t_1+\cdots+r_{|\mathbf{n}|}t_{|\mathbf{n}|}\right)}\right\rangle,
\label{eq:chat}
\end{equation}
with $|\mathbf{n}|=n_1+\cdots+n_s$. The combinatorial factors in Eq. 
\eqref{eq:A} arise from the symmetry in the arguments of the kernels.
Notice that the definition \eqref{eq:chat} leads to
\begin{equation}
A(\mathbf{k},\mathbf{l})=\overline{A(\mathbf{l},\mathbf{k})}
\label{eq:complex}
\end{equation}

Now, making use of the time-shift invariance \eqref{eq:ctshift} in Eq. 
\eqref{eq:chat} implies that $A(\mathbf{k},\mathbf{l})=0$ whenever
$\mathbf{q}\cdot (\mathbf{k}-\mathbf{l}) \ne 0$. Therefore, the only indexes
$\mathbf{k},\mathbf{l} \in\NN_0^s$ in Eq. \eqref{eq:Gammaf} that can contribute to
$\Gamma[\mathbf{f}]$ are those whose difference is a solution of the
Diophantine equation $\mathbf{q}\cdot\mathbf{x}=0$. The set of solutions of
this equation $\mathcal{D}$ can be decomposed as
$\mathcal{D}=\{\mathbf{0}\}\cup\mathcal{D}_+\cup (-\mathcal{D}_+)$. Now, for
every $\mathbf{x}\in\mathcal{D}_+$ let us define $\mathbf{m}=(m_1,\dots,m_s)$
such that
\begin{equation}
m_j=
\begin{cases}
l_j & \text{if $x_j\geqslant 0$,} \\
k_j & \text{if $x_j<0$.} 
\end{cases}
\end{equation}
Thus, if $x_j\geqslant 0$, we can set $l_j=m_j$ and $k_j=m_j+x_j$, whereas if
$x_j<0$, we can set $k_j=m_j$ and $l_j=m_j-x_j$. Denoting $B(\mathbf{m},
\mathbf{x})=A(\mathbf{k},\mathbf{l})$, Eq. \eqref{eq:Gammaf}
becomes
\begin{equation}
\begin{split}
\Gamma[\mathbf{f}]=&\sum_{\mathbf{m}\in\NN_0^s}\left(\prod_{j=1}^s
\epsilon_j^{2m_j}\right)B(\mathbf{m},\mathbf{0}) \\
&+\sum_{\mathbf{m}\in\NN_0^s}\sum_{\mathbf{x}\in\mathcal{D}_+}
\left(\prod_{j=1}^s\epsilon_j^{2m_j+|x_j|}\right)
B(\mathbf{m},\mathbf{x})e^{i\mathbf{x}\cdot\boldsymbol\phi} \\
&+\sum_{\mathbf{m}\in\NN_0^s}\sum_{\mathbf{x}\in\mathcal{D}_+}
\left(\prod_{j=1}^s\epsilon_j^{2m_j+|x_j|}\right)
B(\mathbf{m},-\mathbf{x})e^{-i\mathbf{x}\cdot\boldsymbol\phi}.
\end{split}
\end{equation}
Taking into account that Eq. \eqref{eq:complex} implies $B(\mathbf{m},-\mathbf{x})
=\overline{B(\mathbf{m},\mathbf{x})}$, if we define
\begin{equation}
C_{\mathbf{x}}(\boldsymbol\epsilon)e^{i\theta_{\mathbf{x}}(\boldsymbol\epsilon)}
=\sum_{\mathbf{m}\in\NN_0^s}\left(\prod_{j=1}^s\epsilon_j^{2m_j}\right)
B(\mathbf{m},\mathbf{x}),
\label{eq:resum}
\end{equation}
with $C_{\mathbf{x}}(\boldsymbol\epsilon)=
C_{-\mathbf{x}}(\boldsymbol\epsilon)\ge 0$ and
$\theta_{\mathbf{x}}(\boldsymbol\epsilon)=
-\theta_{-\mathbf{x}}(\boldsymbol\epsilon)\in\RR$, we finally obtain Eq. 
\eqref{eq:cor1}.

\section{Forcing with three or four harmonics}
\label{app:B}

When $q=p=1$ in Eq. \eqref{eq:4harmonic}, the frequency, amplitude, and phase vectors of the three nonzero
harmonics are $\mathbf{q}=(3,1,2)$, $\boldsymbol\epsilon= (3b/2,b/2,a)$, and
$\boldsymbol\phi=(2\phi_1+\phi_2,2\phi_1-\phi_2,
\phi_1+\phi_2)$. Blankinship's algorithm applied to
$\mathbf{q}\cdot\mathbf{x}=0$ yields $\mathbf{x}=k_1(1,-3,0)+k_2(0,2,-1)=
(k_1,2k_2-3k_1,-k_2)$. Then
$\mathbf{x}\cdot\boldsymbol\phi=(4k_1-3k_2)\vartheta$, where
$\vartheta\equiv\phi_2-\phi_1$. Force reversal imposes $k_2-2k_1$ to be odd,
which means that $k_2$ must be odd. Thus, the expansion of the ratchet velocity
will be
\begin{equation}
\begin{split}
v=& \sum_{\substack{k_2=1\\k_2\text{ odd}}}^{\infty}
\tilde C_{0,k_2}(a,b)(b^2a)^{k_2}
\cos\big(3k_2\vartheta- \theta_{0,k_2}(a,b)\big)\\
&+\sum_{k_1=1}^{\infty}\sum_{\substack{k_2\in\ZZ\\k_2\text{ odd}}}
\tilde C_{k_1,k_2}(a,b)b^{k_1+|2k_2-3k_1|}a^{|k_2|}\\
&\times
\cos\big((4k_1-3k_2)\vartheta+ \theta_{k_1,k_2}(a,b)\big),
\end{split}
\label{eq:v4h3}
\end{equation}
where $\tilde C_{k_1,k_2}(a,b)\equiv 3^{k_1}2^{-k_1-|2k_2-3k_1|}
C_{k_1,k_2}(a,b)$. Hence, the lowest order in this expansion is of the
form \eqref{eq:4h3}.

For the case $q=3$, $p=2$ in Eq. \eqref{eq:4harmonic}, the frequency, amplitude, and phase vectors are
$\mathbf{q}=(8,4,5,1)$, $\boldsymbol\phi=(2\phi_1+\phi_2,
2\phi_1-\phi_2, \phi_1+\phi_2, \phi_1-\phi_2)$, and
$\boldsymbol\epsilon=(4b,2b,5a/2,a/2)$, and the solution of
$\mathbf{q}\cdot\mathbf{x}=0$ is $\mathbf{x}=
k_1(1,0,0,-8)+k_2(0,1,0,-4)+k_3(0,0,1,-5)=(k_1,k_2,k_3,-8k_1-4k_2-5k_3)$. Thus,
$\mathbf{x}\cdot\boldsymbol\phi=(3k_1+k_2+2k_3)\vartheta$, where
$\vartheta=3\phi_2-2\phi_1$. Force reversal requires $7k_1+3k_2+4k_3=
k_1+k_2+2(3k_1+k_2+2k_3)$ to be odd; in other words, $k_1+k_2$ must be odd.
Hence,
\begin{equation}
\begin{split}
v= & \sum_{\substack{k_2=1\\k_2\text{ odd}}}^{\infty}\sum_{k_3\in\ZZ}
\tilde C_{0,k_2,k_3}(a,b)a^{|k_3|+|4k_2+5k_3|}b^{k_2} \\
&\times
\cos\big((k_2+2k_3)\vartheta+\theta_{0,k_2,k_3}(a,b)\big)\\
&+\sum_{k_1=1}^{\infty}
\sum_{\substack{k_2\in\ZZ\\k_1+k_2\text{ odd}}}^{\infty}\sum_{k_3\in\ZZ}
\tilde C_{k_1,k_2,k_3}(a,b)a^{|k_3|+|8k_1+4k_2+5k_3|} \\
&\times
b^{k_1+|k_2|}\cos\big((3k_1+k_2+2k_3)\vartheta
+\theta_{k_1,k_2,k_3}(a,b)\big),
\end{split}
\label{eq:v4h4}
\end{equation}
where  $\tilde C_{\mathbf{k}}(a,b)\equiv
5^{|k_3|}2^{2k_1+|k_2|-|k_3|-|8k_1+4k_2+5k_3|}C_{\mathbf{k}}(a,b)$.
The lowest order of this expansion is
\begin{equation}\label{blasteq}
\begin{split}
v= &\tilde C_{0,1,-1}(0,0)ba^2
\cos\big(\vartheta-\theta_{0,1,-1}(0,0)\big)\\
&+\tilde C_{1,-2,0}(0,0)b^3
\cos\big(\vartheta+\theta_{1,-2,0}(0,0)\big)+E_5(a,b),
\end{split}
\end{equation}
where $E_5(a,b)$ contains terms of fifth order in $a$ and $b$. This
expression can be rewritten as in Eq. \eqref{eq:phaseamplitude} by defining
\begin{equation}\label{lasteq}
\begin{split}
A(a,b)e^{i\psi(a,b)}\equiv &\tilde C_{0,1,-1}(0,0)ba^2
e^{-i \theta_{0,1,-1}(0,0)} \\
&+\tilde C_{1,-2,0}(0,0)b^3e^{i \theta_{1,-2,0}(0,0)}.
\end{split}
\end{equation}

\begin{acknowledgments}
We thank Franck Celestini and Ferruccio Renzoni for providing us their
experimental data.  We acknowledge financial support through grants
MTM2012-36732-C03-03 (R.A.N.), FIS2011-24540 (N.R.Q.), PRODIEVO and
Complexity-Net RESINEE (J.A.C.), from Ministerio de Econom\'{\i}a y
Competitividad (Spain), grants FQM262 (R.A.N.), FQM207 (N.R.Q.), and FQM-7276,
P09-FQM-4643 (N.R.Q., R.A.N.), from Junta de Andaluc\'{\i}a (Spain), project
MODELICO-CM (J.A.C.), from Comunidad de Madrid (Spain), and a grant from the
Humboldt Foundation through a Research Fellowship for Experienced Researchers SPA
1146358 STP (N.R.Q.).
\end{acknowledgments}

\end{document}